# Broadband non-reciprocal transmission of sound with invariant frequency

Zhong-ming Gu[1,2], Jie Hu[1,3], Bin Liang[1,2,*], Xin-ye Zou[1] and Jian-chun Cheng[1,2*]

[1]*Key Laboratory of Modern Acoustics, MOE, Institute of Acoustics, Department of Physics, Nanjing University, Nanjing* 210093, *P. R. China*

[2]*Collaborative Innovation Center of Advanced Microstructures, Nanjing University, Nanjing* 210093, *P. R. China*

[3]*Department of Information Science and Technology, Nanjing Forest University, Nanjing,* 210037, *P. R. China*

\* Correspondence and requests for materials should be addressed: B. L. (liangbin@nju.edu.cn) or J. C. C. (jccheng@nju.edu.cn).

**Abstract**

The emergence of "acoustic diode" (AD) capable of rectifying acoustic wave like electrical diodes do to electricity has been believed to be able to offer unconventional manipulation on sound, e.g., to isolate the wrong-way reflection, and therefore have great potential in various important scenarios such as medical ultrasound applications. However, the existing ADs have always been suffering from the problem that the transmitted wave must have either doubled frequency or deviated direction, lacking the most crucial features for achieving such expectations in practice. Here we design and experimentally demonstrate a broadband yet compact non-reciprocal device with hitherto inaccessible functionality of maintaining the original frequency and high forward transmission while virtually blocking the backscattered wave, which is close to what a perfect AD is expected to provide and is promising to play the essential role in realistic acoustic systems like electric diodes do in electrical circuits. Such an extreme ability comes from inherently distinct mechanism based on the exploration of the acoustic characteristics in complex domain, in comparison to the previous designs that only utilize the real part of acoustical parameters. Furthermore, our design enables improving the sensitivity and the robustness of device simultaneously by tailoring an individual structural parameter, which can be regarded as the unique advantage over its electrical or thermal counterparts. We envision our design will take a significant step towards the realization of applicable acoustic one-way devices with potential applications in many scenarios, and inspire the research of non-reciprocal wave manipulation in other fields like electromagnetics.

One-way manipulation of acoustic waves is highly desirable in a great variety of scenarios, but has long been challenging due to the restriction of the well-known reciprocity principle[1-5]. Our realization of an "acoustic diode" (AD) has for the first time broken through this barrier and enabled rectification of acoustic waves by coupling a nonlinear medium with acoustic superlattices[1, 2]. Recently, Alù and colleagues have realized acoustic isolation by using external fluid flows to break space-time symmetry[3]. Popa *et al.* have proposed an active acoustic metamaterials to achieve unidirectional transmission with compact structure[4]. These pioneering works of ADs has attracted many attentions which are believed to offer the potential to revolutionize the existing acoustic techniques in different important fields, e.g., by helping improving the sharpness and resolution of ultrasound images via reduction of wrong-way backscattered waves. In reality, however, this expectation is far beyond reach in respect that in such ADs the transmitted acoustic wave along the "positive" direction has shifted frequency or attenuated amplitude as compared to the incident wave. As a result, such AD prototypes cannot be coupled with other acoustic devices and play the crucial role like its electrical counterpart does in electrical circuits. The emergence of nonlinear ADs has also been followed by considerable efforts dedicated to the pursuit of linear acoustic one-way devices[6-9]. Despite the significantly improved performances of the resulting linear devices including high efficiency, broad bandwidth, and, esp. invariant frequency during transmission, they cannot be regarded as practical ADs since the reciprocity principle still holds in such systems due to their linear nature[10]. In other words, the one-way effect can only be realized for incident wave with particular wavefront incident along particular directions, e.g., normally incident plane wave from two opposite sides. So far, a perfect AD with the potential to really revolutionize the current techniques used in current practical applications, as we have once expected such kind of device to do, still remains challenging.

In this article, we have made a theoretical and experimental attempt to address this difficulty by exploring the potential of acoustic materials in the complex domain, beyond the restriction of the previous designs that only make use of the real part of acoustical parameters. Our proposed model has a sandwich structure that consists of a pair of acoustic gain and lossy media and another acoustical nonlinear material inserted in between. The introduction of gain and lossy effects enables asymmetric manipulation of amplitudes for incident waves from two opposite directions. On other hand, unlike in the existing non-reciprocal designs where the nonlinearity effect is used only for the sake of acoustic frequency conversion[1, 4, 5], we employ the nonlinear material to provide pressure-dependent response to the incident wave, apart from its fundamental necessity for breaking the reciprocal principle[11]. Theoretical analysis reveals that by suitably tuning the structural parameters, the resulting devices, formed by coupling the nonlinear material with the gain/loss pair, could therefore fulfill the functionality a perfect AD is expected to possess that allows acoustic waves to pass along a given direction with a near-unity efficiency and invariant frequency while blocking its transmission along the opposite direction. An experimental implementation by active circuits is also presented, and the results verify our scheme and show that the sample device is able to work as predicted in a broad band despite its compact configuration. With the capability of giving rise to non-reciprocal acoustic transmission while preserving the original frequency and the flexibility of further enhancing the sensitivity and robustness simultaneously, our proposed structure may take a significant step towards the realization of applicable acoustic one-way devices with potential applications in many scenarios.

**Results**

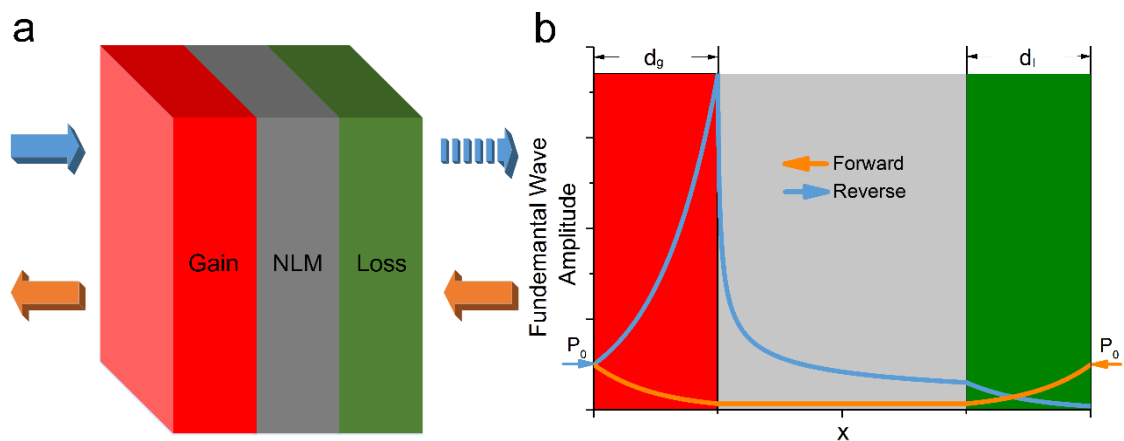

**Figure 1| Design and behavior of the acoustic non-reciprocal device with invariant frequency and near-unity forward transmission.** (a) Schematic of the configuration of the broadband non-reciprocal device. (b) The spatial intensity distribution of the waves coming from the two opposite directions, showing the underlying mechanism responsible for the frequency-invariant rectification phenomenon. The forward (Orange arrow) and the reverse (Blue arrow) directions can be generally defined as the propagating directions of the wave incident from the side of loss and gain materials, respectively. The dotted-line arrow indicates that the transmission along

this direction virtually vanishes.

Figure 1 schematically shows how the desired non-reciprocal transmission can be realized in the proposed structure comprising an acoustic nonlinear material and a pair of gain and lossy materials with well-tuned complex parameters. The sandwich configuration of our design is displayed in Fig. 1(a). In general, the forward direction of such one-way devices can be defined as the propagating direction of wave incident from the side of the lossy material (from right to left for this particular case, as the orange arrow shows), and the opposite direction can be regarded as the reverse direction (blue arrow). As one of the most important characteristics of nonlinear medium filling the middle region of the whole system, it yields quite different responses when the amplitude of the incident wave varies. When the pressure amplitude of the incident wave is at a sufficiently low level, the nonlinear medium will degenerate to a quasi-linear system and will respond linearly to such a small-amplitude wave. If the viscosity effect is negligible in this medium, the incident wave can pass through it without any change in the amplitude or frequency. As we gradually increase the strength of the input signal, however, the medium will exhibit stronger and stronger nonlinearity effect, characterized by the enhancement of the amount of acoustic energy transferred into second and higher harmonic waves, which implies that the transmission of the remained fundamental wave is amplitude-dependent and will decrease as the input acoustic wave propagates. This offers the possibility to give rise to asymmetric transmissions for acoustic wave incident from two opposite sides if the amplitudes of signals from left and right sides can be harnessed differently via controlled amplification and attenuation processes. Equivalently, the manipulation of pressure amplitude can be realized by tailoring the effective refractive index in the complex domain. The recent advance in acoustic materials has extended the concept of acoustic manipulation from passive control[12-16] to active control[4, 17-19], which gives rise to the possibility to control the complex acoustical parameters of materials. For a plane acoustic wave propagating in a homogeneous medium, the acoustic pressure can be expressed as: $p = p_0 e^{j(\omega t - k_0 x)}$. Here, $\omega$ and $k_0$ are the angular frequency and the wave number in the background medium, respectively. For material with a complex refractive index $n = n_r + i n_i$, this relationship can be rewritten as $p = p_0 e^{k_0 n_i x} e^{j(\omega t - k_0 n_r x)}$. If the imaginary part of effective refractive index is positive ($n_i > 0$), the amplitude of acoustic waves will be exponentially amplified as the propagating distance increases, and this kind of material can be referred to as a gain medium whose practical implementation will be discussed later. For the so-called lossy medium characterized by an effective refractive index with a negative imaginary part ($n_i < 0$), the amplitude of acoustic waves will undergo an exponential attenuation as it propagates, and this kind of material can be marked as lossy medium. When we couple the nonlinear medium with a pair of gain and lossy medium as shown in Fig. 1(a), an asymmetric transmission can be expected since the wave incident from the two sides will enter the nonlinear medium with quite different pressure amplitude.

A large gap still needs to be filled, however, between such an asymmetric transmission and the desired one-way effect. For the purpose of achieving a perfect non-reciprocal transmission for which the forward transmitted waves will keep the original amplitude and frequency characteristics, it is vital to delicately tailor the structural parameters in the system to meet two crucial requirements: the gain and lossy must be sufficiently high and, meanwhile, satisfy the complementary condition[20]. For acquiring a clearer insight into the underlying physics in our design, we will analyze the behaviors of the incident wave travelling along the forward and the reverse directions respectively, which can be predicted qualitatively by the results in Fig. 1(b) displaying the spatial distributions of the acoustic energy within the structure for these two particular cases. The subscripts G and L represent the gain and lossy media in the following discussion, respectively. When the gain and loss of the system are high enough and in balance, the acoustic wave incident from the right side impinge on the lossy medium first within which the amplitude of the incident waves will be attenuated to a considerably low value and can be expressed as $p_0 e^{k_0 n_L d_L}$. Due to the quasi-linear nature of the nonlinear medium in this case, the wave will leave this region with preserved frequency and attenuated amplitude. After that, the wave will enter the gain materials where the amplitude can be amplified as $p_0 e^{k_0 (n_L d_L + n_G d_G)}$. Then, it will be restored to the original value because the whole structure has a balanced gain and loss, i.e., $n_L d_L + n_G d_G = 0$, as shown by the orange line in Fig. 1(b). However, when the acoustic wave comes from the reverse side, as the blue line in Fig.

1(b) indicates, it will enter the gain medium first and then leave it with amplitude amplified to a remarkably high value $p_0 e^{k_0 n_G d_G}$, which means the nonlinear conversion from the fundamental waves to high orders harmonic waves will be extremely strong within the nonlinear medium. This obviously results in significantly diminished amplitude of the fundamental frequency and the value $p_0'$ will be much smaller than $p_0 e^{k_0 n_L d_L}$. When the resulting wave passes through the lossy medium, the amplitude of the fundamental wave will undergo a second decrease, which can be expressed as $p_0' e^{k_0 n_L d_L}$. As a result, the total transmission along this direction may become negligible if the conversion rate and loss are high enough. This can be understood as the essential mechanism of the designed non-reciprocal device.

It is worth stressing, however, despite the seemingly similarity between the sandwich configuration illustrated in Fig. 1(a) and the previous ones that rely on nonlinear materials as well, there exists essential difference between the roles of the nonlinear materials in these systems, which consequently results in the distinct advantages of our design in comparison to the existing ones. In the existing designs of non-reciprocal acoustic systems, the acoustic nonlinear medium merely work as a frequency shifter, and the transmitted energy needs to be conveyed in the forms of second harmonic waves, which largely restricts the forward transmission efficiency, let alone the fact that the resulting devices unable to be cascaded with other acoustic devices since the output signal unavoidably has biased frequency. To acquire a decent transmission ratio, the incident wave needs to have very high amplitude. Even if these devices are excited by the large amplitude signal, the transmission efficiency cannot be enhanced to near-unity level due to the fact that the energy of fundamental wave will continue to be converted to higher harmonics. Such a nonlinear saturation effect will hinder the forward transmission from further increasing no matter how strong the input signal is. In the current design, contrarily, the device is designed to exhibit a quasi-linear response to the incident wave along the forward direction, and the transmitted signal is the fundamental wave rather than second harmonic wave. This is responsible for the crucial feature a perfect AD is desired to have, i.e., the original frequency and amplitude of the incident wave can remain almost unaffected. In the reverse direction, the pre-amplification in the gain medium can significantly reduce the requirement on the amplitude of the incident wave to yield a certain amount of nonlinear conversion. Furthermore, since one no longer relies on the proportion of the energy of second harmonic wave, the fundamental component of the incident wave will be further attenuated despite the existence of the above saturation effect, which will eventually lead to the blockage of transmission along this direction.

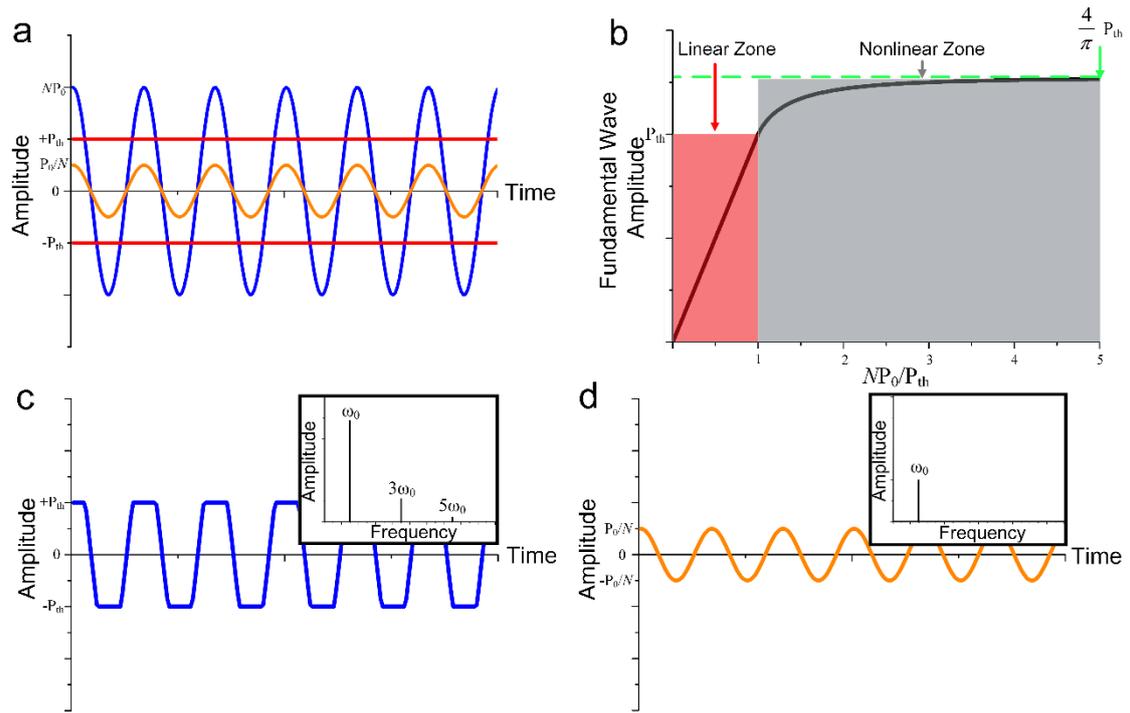

**Figure 2 | Time-domain and frequency-domain waveforms of the input and output signals of the nonlinear medium.** (a) Time-domain input waveform along the forward and reverse directions after passing through the lossy and gain media they meet first. (b) The amplitude dependence of the nonlinear medium. The time-domain signal after passing through the nonlinear medium along the reverse direction (c) and forward (d) directions respectively. The insets are the corresponding spectrum components. Along the reverse direction, the wave has its peak being clipped off when passing through the nonlinear medium, and the amplitude of transmitted fundamental wave is then largely diminished. The wave travelling along the positive direction is not affected by the nonlinear medium due to its small amplitude.

For a clear physics picture, we will inspect the time-domain waveform of the wave entering and leaving the nonlinear medium, the key part providing pressure-dependent property that is crucial for achieving the nonreciprocal transmission, and analyze the variation in the amplitude of spectrum components as the wave travels through this part along the two opposite directions. As a typical nonlinear circuit, a limiter circuit allows signals below a specified input power to pass unaffected while attenuating the peaks of stronger signals that exceed this input power. Here, we endow the nonlinear medium with the property of limiter circuit that has a 'turn on' value to distinguish the linear zone and nonlinear zone. Then, we will have three typical signals in this system, as shown in Fig. 2(a). The blue (orange) sinusoid represents the amplified (attenuated) wave which can be regarded as the output signal of the gain (lossy) material and will pass through the nonlinear medium along reverse (forward) direction. And the red line represents the specified value of the limiter, which keeps a constant over time. Here, $p_0$, $p_{th}$ and $N$ are the original amplitude of the incident waves, the threshold value and the amplified factor. With the changing of the amplified factor from small to large, it is obvious that the nonlinear part will have a different response to the incident waves, as shown in Fig. 2(b). At first, the amplified factor is very small and the amplified value is smaller than the threshold value, then the limiter will work on the linear zone, as depicted as red zone, which means the transmitted waves will have the same frequency characteristic as the incident waves. However, when the amplified value exceeds the specified value, the limiter will have a nonlinear response, as depicted as grey zone. The high orders harmonic waves will be generated and the amplitude of the fundamental wave will approach saturation. Besides, when the amplified value is large enough, the distorted sinusoid waves will be a very close approximation to the square waves, which means the saturation point $\frac{4}{\pi}p_{th}$ can be predicted by the Fourier transform of the square waves, as depicted as green dashed lines. Apparently, in this design, it is desired that the amplified amplitude is much larger than the specified value of the limiter while the attenuated amplitude is quite below it. Then the nonlinear part can be used to modulate the acoustic wave incident from the two sides to create the non-reciprocal transmission. Figure 2(c) shows the amplified waves modulated by the limiter in time domain and the inset is the frequency spectrum. It is clear that along the reverse direction, the wave has to have its peak being clipped off when passing through the nonlinear medium, which leads to the generation of the high orders harmonic waves and significant reduction in the amplitude of transmitted fundamental wave. The wave will then enter the lossy medium to undergo the second attenuation. On the other hand, Fig. 2(d) shows the time-domain waveform of the attenuated waves after passing through the limiter and the inset is the corresponding spectrum components. The wave travelling along the forward direction is not affected by the nonlinear medium due to the fact that the amplitude of such attenuated waves is lower than the turn-on value, the limiter works in the linear zone, which means the transmitted wave will have an invariant frequency. The wave along this direction will then have its amplitude restored to the initial value after propagating through the gain medium.

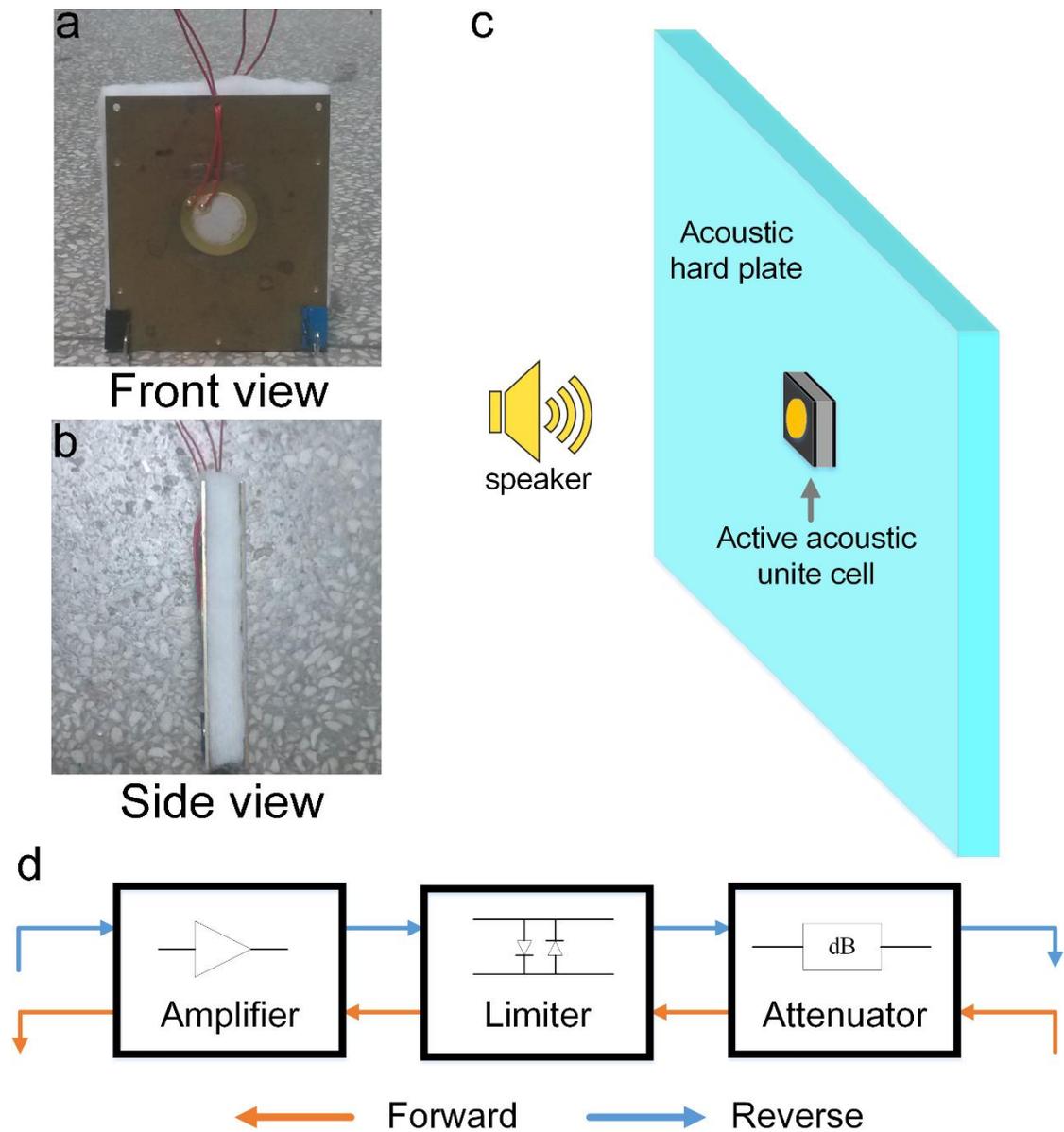

**Figure 3 | Experimental set-up.** Photograph of basic unit cell is shown in (a) (Front view) and (b) (Side view). (c) Schematic of the experimental environment. (d) The block diagram of nonlinear electronic circuit.

For verifying the effectiveness of our scheme, we have fabricated a prototype of the designed device and measured its non-reciprocal performance experimentally. For natural acoustic materials, the refractive index should have a negative imaginary part, indicating that the material is lossy with inherent damping. In contrary, acoustic gain materials with a positive imaginary part of the effective refractive index are usually unavailable in nature, which have recently attracted considerable attentions[17, 18, 21, 22]. Cummer and coworkers have achieved the acoustic gain effect by using active acoustic unite cell, which has inspired many potential applications[4, 21]. Compared to the conventional acoustic nonlinear medium, on the other hand, the nonlinear behavior generated by the electronic circuit will be more robust. Therefore we employ electronic circuits in the current study as a specific implementation. We need to emphasize, however, this is only for the sake of an unambiguous demonstration of the effectiveness of our design that is in fact general and may be implemented by any practical acoustical systems satisfying the above-mentioned requirements on the structural parameters. The front view and side view of the fabricated unit cell which has a deep subwavelength scale have been shown in Figs. 3(a) and (b). As seen in the figures, the unit cell has a symmetry structure composed of five layers. For the sake of unambiguously identifying the amount of the transmitted acoustic energy, we fill the middle region of the structure with sound-absorbing foam to absorb the unwanted transmitted waves and isolate the vibration, and couple the sound-absorbing layer to two brass plates, which can acoustically treated as a rigid boundary for airborne sound, to block the waves. Two 35-mm-diameter piezoelectric (PZT) ceramic films, which have been widely used in the previous designs of acoustic active unit cell and proven to be quite good candidate for acoustic gain material, have been attached to the two brass plates. Here, we employ it as sensing and driving transducers since it has the ability of reciprocal conversion between acoustic energy and electric energy[23]. The experiment set-up has been shown in Fig. 3(c). The unit cell we designed has been fixed in the middle of an acoustic hard plate which is used here to avoid the unnecessary diffraction. The incident wave is excited by a speaker in one side of the plate and the transmitted signal is measured in the other side. Two PZT ceramic films are connected to the electronic circuit whose block diagram is presented in Fig. 3(d). The amplifier circuit is based on the design of LM386 audio amplifier and has an amplification of 20[24]. The limiter circuit mainly consists of an antiparallel diode pair. And some dissipative conductors are arranged to implement the attenuator part. Furthermore, this device can be conveniently tuned to work for incident waves with different amplitudes by simply adjusting the parameters in the circuit.

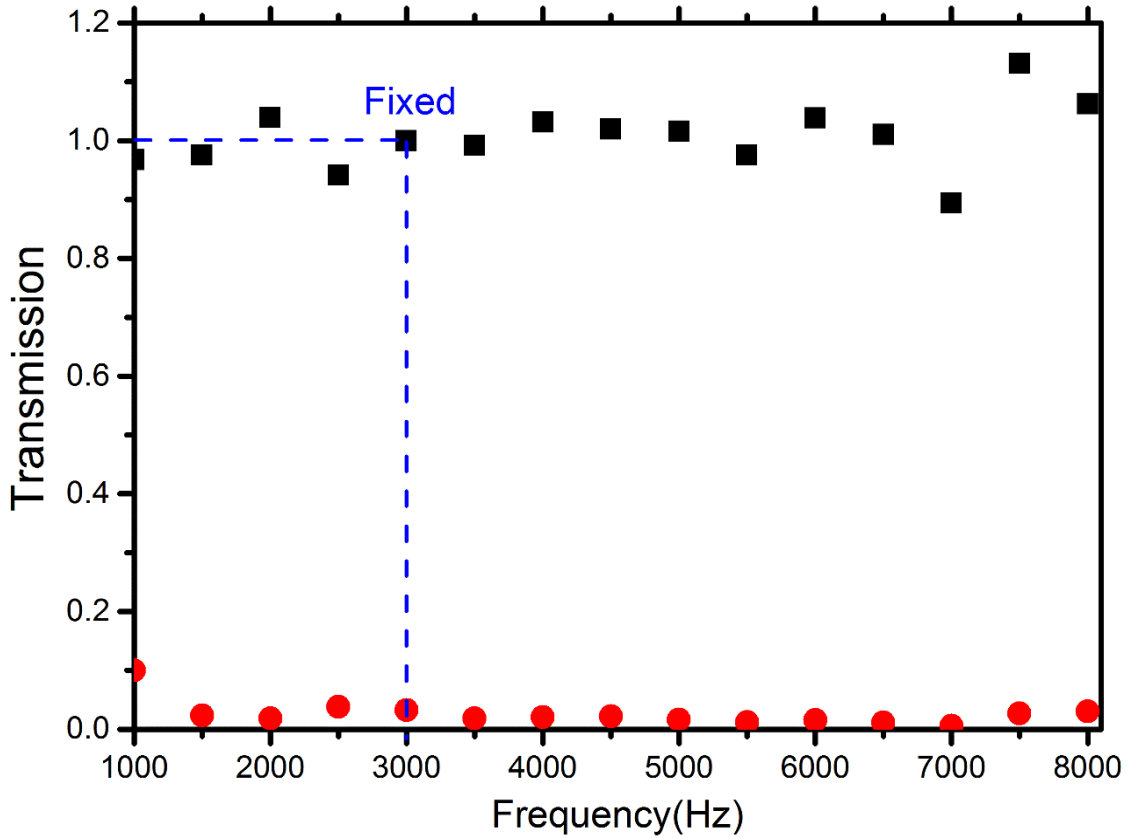

**Figure 4 | The transmission measurements**. The experimental results of the transmissions along the forward (black squares) and reverse directions (red dots) are plotted for comparison. The blue dashed line indicates the location of the working point, chosen as 3kHz in the current design, for which the resulting device is desired to yield the optimal performance. The measurements were repeated 3 times and the results are virtually identical each time.

We have measured the performance of the prototype within a broad frequency band ranging from 1kHz to 8kHz. The speaker is controlled by a signal generator and launches sinusoid signal for which the frequency is swept with a step of 500Hz. To characterize the non-reciprocal transmission efficiency, we measure the amplitude of incident signals and transmitted signals along the two opposite directions, and evaluate the discrepancy between the transmissions of forward and reverse waves. For a quantitative estimation, we define the transmission along the forward direction as $T_f$ and the transmission along the reverse direction as $T_r$ respectively. The measurement results have been plotted in Fig. 4 in which $T_f$ and $T_r$ are represented by black square scatter and red circle scatter, respectively. In the measurement, we have adjusted the system parameter delicately to guarantee an optimal performance at a particular frequency of the device, chosen as 3kHz in the current design, which means that the forward transmission is virtually unity as the driving frequency is exactly at this frequency. When the incident wave of the same frequency travels along the reverse direction, the transmission will be significantly diminished, as predicted from the fundamental mechanism in the structure we propose. It is noteworthy, however, although the working point is set at this individual frequency, non-reciprocal transmission property of the prototype persists within an ultra-broad band except for slight fluctuations of the values of the forward transmissions around $T_f = 1$ which may be caused by the dynamic impedance mismatch in the system.

**Discussion**

Realization of a perfect AD has long been challenging, to which considerable efforts have been and continue to be devoted. In the previous works, the schemes are mainly the combination of nonlinear medium and some artificial structures, like acoustic superlattices or Helmholtz cavities. The former is used to shift the frequency and the latter is used as a frequency filter. Besides, by introducing the external fluid flow, the acoustic circle isolation can be achieved. However, it is inevitable that the transmitted waves have a shifted frequency as compared to the original incident wave. This results in the problem that the forward transmission is limited by the efficiency of nonlinear conversion and the resulting devices cannot be cascaded like their electrical counterparts. By exploiting the characteristic of nonlinear medium that it has a nonlinear response to the input signals with different amplitude, our proposed structure consists of acoustic nonlinear medium, acoustic gain and lossy materials to realize the nonreciprocal transmission. Compared the previous designs, the transmitted waves have an invariant frequency exactly the same as the incident waves and the amplitude can be tuned to unity, which are very close to what a perfect AD is expected to offer. Furthermore, it is observable from Fig. 2 that by adjusting the parameter of the nonlinear medium, we can further reduce its turn-on value representing the transition from the linear regime to the nonlinear regime, which may endow the resulting device with an intriguing feature. By reducing this individual parameter, it is possible to decrease the threshold value for the amplitude of incident wave to "conduct" along the forward direction and, meanwhile, make it harder for the structure to allow the reverse wave with extremely large amplitude to pass in a similar manner to the breakdown effect in electrical diodes. In other words, this offers the potential to enhancing the sensitivity and robustness of our one-way device simultaneously, which is of paramount significance for the application of such devices in practice. Consequently, we anticipate our design, with the capability of realizing broadband non-reciprocal transmission and the flexibility of being tailored to further improving the performance, will make a significant step forward in the pursuit of a perfect AD and have potential applications in a great variety of scenarios.

**Acknowledgement**

This work was supported by the National Basic Research Program of China (973 Program) (Grant Nos. 2010CB327803 and 2012CB921504), National Natural Science Foundation of China (Grant Nos. 11174138, 11174139, 11222442, 81127901, and 11274168), NCET-12-0254, and A Project Funded by the Priority Academic Program Development of Jiangsu Higher Education Institutions.

**Author contributions**

Z. M. G., B. L., X. Y. Z. performed analytical and numerical computations. Z. M. G., B. L., J. H. conducted the circuit design and the experiments. B. L. and J. C. C. conceived and supervised the study. Z. M. G., J. H., B. L. and J. C. C wrote the article. All authors contributed to the discussions.